\documentclass{ws-procs9x6}

\usepackage{graphicx,amssymb}

\def\bea{\begin{eqnarray}}
\def\eea{\end{eqnarray}}
\def\be{\begin{equation}}
\def\ee{\end{equation}}
\newcommand{\ub}[1]{\underline{#1}}

\def\eqref#1{(\ref{#1})}

\begin{document}

\title{%
SUPERSYMMETRIC TWO-DIMENSIONAL QCD AT FINITE TEMPERATURE%
\footnote{
To appear in the proceedings of the seventh workshop on 
\uppercase{C}ontinuous \uppercase{A}dvances in \uppercase{QCD},
\uppercase{M}inneapolis, \uppercase{M}innesota, 
\uppercase{M}ay 11-14, 2006.}
}

\author{J.R. HILLER}

\address{Department of Physics \\
University of Minnesota-Duluth \\
Duluth, MN 55812 USA \\
E-mail: jhiller@d.umn.edu}

\begin{abstract}
Light-cone coordinates and 
supersymmetric discrete light-cone quantization are used
to analyze the thermodynamics
of two-dimensional supersymmetric quantum chromodynamics with
a Chern--Simons term in the large-$N_c$ approximation.
This requires estimation of the entire spectrum of the theory, which is
done with a new algorithm based on Lanczos iterations.
Although this work is still in progress, some preliminary results
are presented.
\end{abstract}

\keywords{supersymmetry, quantum chromodynamics, finite temperature, 
density of states}

\bodymatter

\section{Introduction}

Recent work\cite{FiniteTemp} has shown that thermodynamic properties
can be computed for large-$N_c$ supersymmetric theories.  The approach
is based on light-cone coordinates\cite{Dirac} and the numerical technique
of supersymmetric discrete light-cone quantization (SDLCQ).\cite{Sakai,SDLCQreview}
Here we consider two-dimensional supersymmetric quantum chromodynamics with
a Chern--Simons term (SQCD-CS),\cite{SQCD-CS} dimensionally reduced from
three dimensions.

Light-cone coordinates\cite{Dirac} are defined by the time variable, $x^+=(t+z)/\sqrt{2}$,
and spatial components, $\ub{x}=(x^-,\vec{x}_\perp)$, where
$x^-\equiv (t-z)/\sqrt{2}$ and $\vec{x}_\perp=(x,y)$.  The light-cone energy and
momentum are given by $p^-=(E-p_z)/\sqrt{2}$ and
$\ub{p}=(p^+\equiv (E+p_z)/\sqrt{2},\,\vec{p}_\perp=(p_x,p_y))$, respectively.

For field theories quantized in terms of these coordinates, the
standard numerical technique is discrete light-cone quantization 
(DLCQ).\cite{PauliBrodsky,DLCQreview}  Space is restricted to a
light-cone box $-L<x^-<L$, $-L_\perp<x,y<L_\perp$ with periodic 
boundary conditions.  Momentum is then discretized as
$p_i^+\rightarrow\frac{\pi}{L}n_i$,
 ${\bf p}_{i\perp}\rightarrow
     (\frac{\pi}{L_\perp}n_{ix},\frac{\pi}{L_\perp}n_{iy})$
     with $n_i$, $n_{ix}$ and $n_{iy}$ all integers.
The limit $L\rightarrow\infty$ is exchanged for a limit
in terms of the integer (harmonic) resolution\cite{PauliBrodsky}
$ K\equiv\frac{L}{\pi}P^+$ for fixed total momentum $P^+$.
Because the $n_i$ are positive, the number of particles is
limited to no more than $K$. Integrals are replaced by discrete sums.

SDLCQ\cite{Sakai,SDLCQreview} is
a special form of DLCQ that preserves at least part of the
supersymmetry algebra
\be
\{Q^+,Q^+\}=2\sqrt{2}P^+, \;\;
\{Q^-,Q^-\}=2\sqrt{2}P^-, \;\;
\{Q^+,Q^-\}=-4P_\perp.
\ee
Instead of discretizing the Hamiltonian $P^-$ directly,
the supercharge $Q^-$ is discretized, and $P^-$ is
computed from the algebra as
\be
P_{\rm SDLCQ}^-=\frac{1}{2\sqrt{2}}\left\{Q^-,Q^-\right\}
                         \neq P_{\rm DLCQ}^-.
\ee
For ordinary DLCQ, one recovers supersymmetry only
in the infinite resolution limit.

After a brief summary of the SQCD-CS theory, we show how
thermodynamic quantities can be constructed from the 
partition function.  This requires knowledge of the spectrum
of the theory, which we obtain numerically with an iterative
Lanczos algorithm.  Some preliminary results are presented
and future work discussed.

\section{Supersymmetric QCD}

We consider a dimensional reduction from 2+1 to 1+1 dimensions
of ${\cal N}=1$ supersymmetric quantum chromodynamics with a 
Chern--Simons term (SQCD-CS).\cite{SQCD-CS}  The action is
\bea\label{action}
S&=&\int d^3x\mbox{Tr}\left\{-\frac{1}{4}F_{\mu\nu}F^{\mu\nu}
+D_\mu \xi^\dagger D^\mu \xi
+i{\bar\Psi} D_\mu\Gamma^\mu\Psi\right.
\nonumber\\
&&-g\left[{\bar\Psi}\Lambda\xi
+\xi^\dagger{\bar\Lambda}\Psi\right]
+\frac{i}{2}{\bar\Lambda}\Gamma^\mu D_\mu \Lambda \\
&&\left.+\frac{\kappa}{2}\epsilon^{\mu\nu\lambda}
  \left[A_{\mu}\partial_{\nu}A_{\lambda}
           +\frac{2i}{3}gA_\mu A_\nu A_\lambda \right]
+\kappa\bar{\Lambda}\Lambda
\right\},
\nonumber
\eea
where the adjoint fields are the gauge boson $A_\mu$ (gluons)
and a Majorana fermion $\Lambda$ (gluinos) and the fundamental
fields are a Dirac fermion $\Psi$ (quarks) and a complex scalar 
$\xi$ (squarks).
The Chern--Simons coupling,
$\kappa$, has the effect of providing a mass for the adjoint fields.
The covariant derivatives are
\bea
D_\mu\Lambda&=&\partial_\mu\Lambda+ig[A_\mu,\Lambda]\,,\quad
D_\mu\xi=\partial_\mu\xi+igA_\mu\xi\,,  \nonumber \\
D_\mu\Psi&=&\partial_\mu\Psi+igA_\mu\Psi\,.
\eea
The fields transform according to
\bea
&&\delta A_\mu=\frac{i}{2}{\bar\varepsilon}\Gamma_\mu\Lambda\,,\qquad
\delta\Lambda=\frac{1}{4}F_{\mu\nu}\Gamma^{\mu\nu}\varepsilon\,,\nonumber\\
&&\delta\xi=\frac{i}{2}{\bar\varepsilon}\Psi\,,\qquad
\delta\Psi=-\frac{1}{2}\Gamma^\mu\varepsilon D_\mu\xi.
\eea

We reduce to 1+1 dimensions by assuming the fields to be independent of the
transverse coordinate $x$.  We define fermion components by
\be
\Lambda=\left(\lambda,{\tilde\lambda}\right)^T\,,\qquad
\Psi=\left(\psi,{\tilde\psi}\right)^T\,,\qquad
Q=\left(Q^+,Q^-\right)^T.
\ee
There are constraints, which in light-cone gauge ($A^+=0$) are 
written
\bea
\partial_-{\tilde\lambda}&=&-\frac{ig}{\sqrt{2}}
\left([A^2,\lambda]+i\xi\psi^\dagger-i\psi\xi^\dagger\right),\\
\partial_-{\tilde\psi}&=&-\frac{ig}{\sqrt{2}}A^2\psi+
\frac{g}{\sqrt{2}}\lambda\xi -\kappa\lambda/\sqrt{2}\,, \;\; \label{fcurrent} 
\partial^2_-A^-=gJ\,,
\eea
with
\be \label{current}
J\equiv i[A^2,\partial_-A^2]+
\frac{1}{\sqrt{2}}\{\lambda,\lambda\}
+\kappa \partial_- A^2
-ih\partial_-\xi\xi^\dagger+
i\xi\partial_-\xi^\dagger+\sqrt{2}\psi\psi^\dagger\,.
\ee
The reduced supercharge is
\bea
Q^-&=&g\int dx^-\left\{
    2^{3/4}\left({\rm i}[A^2,\partial_-A^2]-\kappa\partial_-A^2
  +\frac{1}{\sqrt{2}}\{\lambda,\lambda\}\right)\frac{1}{\partial_-}\lambda 
  \right.  \nonumber \\
&&-\frac{1}{\sqrt{2}}\left(i\sqrt{2}\xi\partial_
-\xi^\dagger-i\sqrt{2}\partial_-\xi\xi^\dagger+2\psi\psi^\dagger\right)
\frac{1}{\partial_-}\lambda   \\
&&\left.-2\left(\xi^\dagger A^2\psi+\psi^\dagger A^2\xi\right)\right\}\,.
\nonumber
\eea

In the large-$N_c$ approximation, there are only single-trace Fock
states, the mesons
\be
{\bar f}^\dagger_{i_1}(k_1) a^\dagger_{i_1i_2}(k_2)\dots
b^\dagger_{i_ni_{n+1}}(k_{n-1})\dots f^\dagger_{i_p}(k_n)|0\rangle
\ee
and glueballs
\be
{\rm Tr}[a^\dagger_{i_1i_2}(k_1)\dots
b^\dagger_{i_ni_{n+1}}(k_{n})]|0\rangle,
\ee
where
$\bar{f}_i^\dagger$ and $f_i^\dagger$ create fundamental partons and
$a_{ij}^\dagger$ and $b_{ij}^\dagger$ create adjoint partons.  Either
type of state could be a boson or a fermion.

The theory possesses a useful $Z_2$ symmetry\cite{Kutasov}
\be
a_{ij}(k,n^\perp)\rightarrow -a_{ji}(k,n^\perp), \;\;
b_{ij}(k,n^\perp)\rightarrow -b_{ji}(k,n^\perp),
\ee
which further divides the Fock space between states with even and odd numbers 
of gluons.  We then diagonalize in each sector separately.

\section{Finite Temperature}

From the partition function, $Z=e^{-p_0/T}$,
we compute the bosonic free energy
\be
{\cal F}_B=\frac{VT}{\pi}\sum_{n=1}^\infty
\int_{M_n}^\infty dp_0
\frac{p_0}{\sqrt{p_0^2-M_n^2}}
\ln \left( 1- e^{- p_{0}/T }\right).
\ee
and the fermionic free energy
\be
{\cal F}_F= -\frac{VT}{\pi}\sum_{n=1}^\infty
   \int_{M_n}^\infty dp_0
\frac{p_0}{\sqrt{p_0^2-M_n^2}}
 \ln \left( 1+ e^{- p_{0}/T }\right).
\ee
The total free energy, once expanded in logarithms and $p_0$ integrals
are performed, is given by 
\be
{\cal{F}}(T,V)=-\frac{(K-1)\pi}{4}
VT^2-\frac{2VT}{\pi}\sum_{n=1}^{\infty}{\sum_{l=0}^{\infty}}
M_{n}\frac{K_{1}\left((2l+1)\frac{M_{n}}{T}\right)}{(2l+1)}.
\ee
The sum over $l$ is well approximated by the first few terms.
We represent the sum over $n$ as an integral over a density of states
$\rho$:
$\sum_n \rightarrow \int \rho(M) dM $.
The density is approximated by a continuous function,
and the integral $\int dM$ is computed by standard numerical techniques.

\section{Lanczos Algorithm for Density of States}

The discrete density of states is $\rho(M^2)=\sum_n d_n \delta(M^2-M_n^2)$,
where $d_n$ is the degeneracy of the mass eigenvalue $M_n$.
The density can be written in the form of a trace over the
evolution operator $e^{-iP^-x^+}$:
\be
\rho(M^2)=\frac{1}{4\pi P^+}\int_{-\infty}^\infty e^{iM^2x^+/2P^+}
           {\rm Tr}e^{-iP^-x^+} dx^+.
\ee
We approximate the trace as an average over a
random sample of vectors\cite{AlbenHams}
\be
\rho(M^2)\simeq\frac1S\sum_{s=1}^S \rho_s(M^2),
\ee
with $\rho_s$ a local density for a single vector $|s\rangle$,
defined by
\be
\rho_s(M^2)=\frac{1}{4\pi P^+}\int_{-\infty}^\infty e^{iM^2x^+/2P^+}
           \langle s|e^{-iP^-x^+}|s\rangle dx^+.
\ee
The sample vectors $|s\rangle$ can be chosen as random 
phase vectors;\cite{Iitaka} the coefficient
of each Fock state in the basis is a random number of 
modulus one.
  
We approximate the matrix element $\langle s|e^{-iP^-x^+}|s\rangle$
by Lanczos iterations.\cite{JaklicAichhorn}
Let $D$ be the length of $|s\rangle$, and define
$|u_1\rangle=\frac{1}{\sqrt{D}}|s\rangle$ as the
initial Lanczos vector. The matrix element
$\langle u_1|e^{-iP^-x^+}|u_1\rangle$ can be
approximated by the $(1,1)$ element of the 
exponentiation of the Lanczos tridiagonalization
of $P^-$.   Let $P_s^-$ be the tridiagonal Lanczos matrix.
It can be exponentiated by first diagonalizing it:
\be
P_s^-\vec{c}_j^{\,s}=\frac{M_{sj}^2}{2P^+}\vec{c}_j^{\,s},
\ee
such that $P_s^-=U\Lambda U^{-1}$, with $U_{ij}=(c_j^s)_i$
and $\Lambda_{ij}\equiv\delta_{ij}\frac{M_{sj}^2}{2P^+}$.  
The $(1,1)$ element is given by
\be
\left(e^{-iP_s^-x^+}\right)_{11}=\sum_n|(c_n^s)_1|^2e^{-iM_{sn}^2 x^+/2P^+}.
\ee

The local density can now be estimated by
\be
\rho_s(M^2)\simeq \sum_n w_{sn} \delta(M^2-M_{sn}^2),
\ee
where $w_{sn}\equiv D|(c_n^s)_1|^2$ is the weight of each Lanczos eigenvalue.
Only the extreme Lanczos eigenvalues are good approximations to
eigenvalues of the original $P^-$. The other Lanczos eigenvalues 
provide a smeared representation of the full spectrum.

From this density of states, we compute the cumulative distribution function (CDF),
$N(M^2)=\int^{M^2}dM^2 \rho(M^2)$
as an average 
\be
N(M^2)\simeq\frac1S\sum_s N_s(M^2),
\ee
of local CDFs
\be
N_s(M^2)\equiv \int^{M^2} dM^2\rho_s(M^2)\simeq\sum_n w_{sn}\theta(M^2-M_{sn}^2).
\ee

The convergence of the approximation is dependent on the number of
Lanczos iterations per sample, as well as the number $S$ of samples.
Test runs indicate that taking 20 samples is sufficient.  
The number of Lanczos iterations needs to be on the
order of 1000 per sample; using only 100 leaves errors on the order
of 1-2\%.

\section{Preliminary Results}

Some preliminary results are presented in the accompanying figures.
Figure 1 compares the numerical results for the CDF to analytic
results, which can be obtained when the Yang-Mills coupling is 
zero.  The numerical results are quite good, with only one 
noticeable deviation, at large $M^2$, a region where the discrete
spectrum is sparse.  Figure 2 shows the free energy at particular
values of the Yang--Mills coupling.
\begin{figure}[ht]
\centerline{\includegraphics[width=10cm]{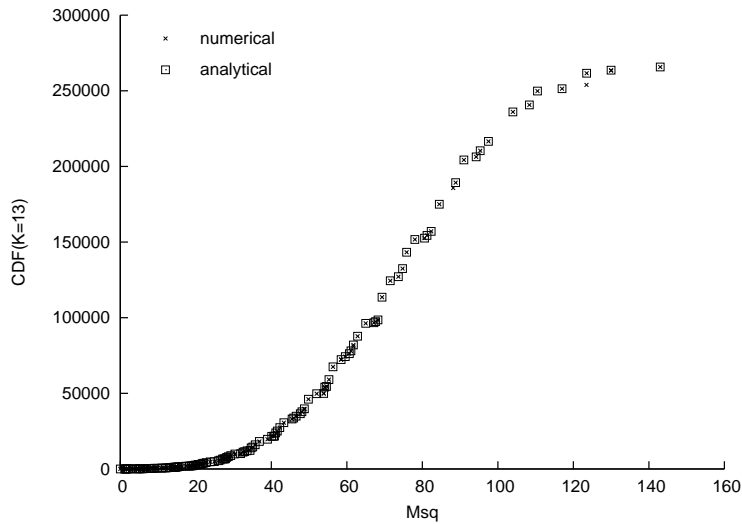}}
\caption{Cumulative distribution function for resolution $K=13$
in the analytically solvable case of zero Yang--Mills coupling.
The numerical and analytic solutions are compared.}
\end{figure}
\begin{figure}[ht]
\centerline{\includegraphics[width=10cm]{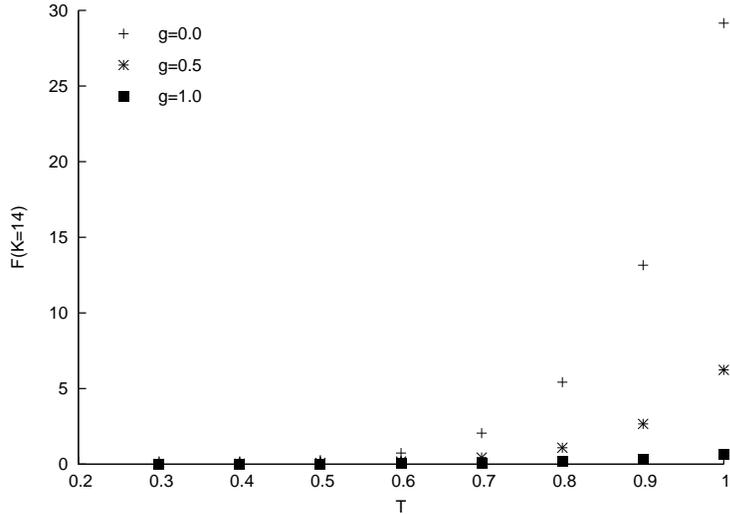}}
\caption{Free energy at fixed Yang--Mills coupling $g$ as a function
of temperature $T$ for resolution $K=14$.}
\end{figure}

\section{Future Work}

Additional work is in progress to complete this study of
finite temperature properties of two-dimensional SQCD.
Beyond this particular effort, one can consider finite-$N_c$
effects, with baryons and mixing of mesons and glueballs,
and the full three-dimensional theory.  As these techniques
mature, analysis of four-dimensional theories can be 
considered.

\section*{Acknowledgments}
The work reported here was done in collaboration with 
S.~Pinsky, Y.~Proestos, N.~Salwen, and U.~Trittmann
and supported in part by the US Department of Energy
and the Minnesota Supercomputing Institute.

\end{document}